\documentclass[smallcondensed]{svjour3}    
\smartqed  
\usepackage{graphicx}
\usepackage[english]{babel}
\usepackage{amssymb}
\usepackage{mathrsfs}
\usepackage{amsmath}
\usepackage{rotate}
\usepackage{vmargin}

\begin{document}
\title{Complex statistics in Hamiltonian barred galaxy models}

\author{Tassos~Bountis \and Thanos~Manos \and Chris~Antonopoulos}

\institute{T.~Bountis \at
              Department of Mathematics and Center for Research and
              Applications of Nonlinear Systems (CRANS), University of Patras,
              GR-$26500$, Rion, Patras, Greece\\
              Tel./Fax: +30-2610-997-381\\
              \email{bountis@math.upatras.gr} \and
              T.~Manos \at
              Department of Mathematics and Center for Research and
              Applications of Nonlinear Systems (CRANS), University of Patras,
              GR-$26500$, Rion, Patras, Greece\\
              \email{thanosm@master.math.upatras.gr} \and
              Ch.~Antonopoulos \at
              Department of Automation and High Performance Computing Systems - Programming \& Algorithms Lab (HPCS Lab), Technological Educational Institute of Messolonghi,
              Nea Ktiria, 30200, Messolonghi, Greece\\
              \email{chantonopoulos@teimes.gr}
              }
\maketitle

\begin{abstract}
We use probability density functions (pdfs) of sums of orbit coordinates, over time intervals of the order of one Hubble time, to distinguish weakly from strongly chaotic orbits in a barred galaxy model. We find that, in the weakly chaotic case, quasi-stationary states arise, whose pdfs are well approximated by $q$-Gaussian functions (with $1<q<3$), while strong chaos is identified by pdfs which quickly tend to Gaussians ($q=1$). Typical examples of weakly chaotic orbits are those that ``stick'' to islands of ordered motion. Their presence in rotating galaxy models has been investigated thoroughly in recent years due of their ability to support galaxy structures for relatively long time scales. In this paper, we demonstrate, on specific orbits of 2 and 3 degree of freedom barred galaxy models, that the proposed statistical approach can distinguish weakly from strongly chaotic motion accurately and efficiently, especially in cases where Lyapunov exponents and other local dynamic indicators appear to be inconclusive.



\end{abstract}

\section{Introduction}\label{intro}

In galactic dynamics, the identification of orbits as ordered (i.e. periodic or quasiperiodic) or chaotic and their respective role in the structure of galaxies has been studied extensively for many years (see \cite{Con_spr} for a recent review). Distinguishing between quasiperiodic and chaotic motion, in general, poses no major difficulty, as there are many numerical approaches presently available by which this distinction can be easily accomplished. What is considerably more complicated is the identification of \textit{weakly} chaotic orbits, particularly when integration times are not long enough. Weak chaos is manifested, for example, in trajectories that ``stick'' for long times on the boundaries of ``islands'' of ordered motion, where the maximal Lyapunov exponent (MLE) is positive but ``small'', before eventually occupying a wider domain of strong chaos with much larger MLEs.

Weakly chaotic orbits spend a significant fraction of their time in confined regimes and do not fill up phase space as ``homogeneously'' as strongly chaotic trajectories, which explore faster and more uniformly their available regions. Several researchers have demonstrated the importance of weakly chaotic phenomena in supporting barred or spiral galaxy features (see e.g. \cite{Athan:2009a,Athan:2009b,Athan:2010,HaKa:2009,KauCo,PAQ1997,Pats:2006,Rom:2006,Rom:2007}). There are also several results in the recent literature showing that strong local instability does not necessarily imply widespread diffusion in phase space \cite{Cachuchoetal:2010,CinGiochap:2008,Gio:2004}. In \cite{ConHa:2008,ConHa:2010} ``stickiness'' was studied thoroughly in 2 degrees of freedom (dof) and was categorized in two main types: (a) ``Stickiness'' around an island of stability and (b) ``stickiness'' in chaos, along the invariant manifolds of unstable periodic orbits. The role of ``sticky'' chaotic orbits and the diffusive behavior in the neighborhood of invariant tori surrounding periodic solutions of the Hamiltonian was studied in \cite{KP11}. The challenging problem, however, is to be able to \emph{detect} weak chaos (as manifested e.g. in ``stickiness'' phenomena), for \emph{short} integration times (not exceeding the age of the universe), before the true character of the dynamics becomes apparent.

In the present paper, we propose a novel approach for tackling this problem based on the \textit{statistics} of the orbits of certain barred galaxy models. In particular, we consider probability density functions (pdfs) of \textit{sums} of their variables, in the spirit of the Central Limit Theorem \cite{Rice1995}. If these pdfs quickly tend to Gaussians, we classify the orbits as strongly chaotic and suggest that their motion is characterized Boltzmann-Gibbs statistical mechanics. If, on the other hand, the pdfs are well approximated by $q$--Gaussian functions, we characterize the orbits as weakly chaotic, implying that they form quasi-stationary states that obey the principles of nonextensive statistical mechanics \cite{ABB_QSS,Tsallisbook2009}

Pdfs of chaotic trajectories of dynamical systems have long been studied by many authors, aiming to understand the transition from deterministic to stochastic dynamics \cite{Arnold1967,Eckmann1985,katok1980,Pesin1976,Ruelle1979,Sinai1972}.
The main question here concerns the existence of an appropriate invariant probability measure, characterizing chaotic phase space regions where solutions generically exhibit exponential divergence of nearby trajectories. If this invariant measure turns out to be a continuous and sufficiently smooth function of the phase space coordinates, one can invoke the Boltzmann-Gibbs microcanonical ensemble and attempt to evaluate all relevant quantities of equilibrium Statistical Mechanics, like partition function, free energy, entropy, etc. of the system.

In such cases, viewing the values of one (or a linear combination) of components of a chaotic solution at discrete times $t_n, n=1,\ldots,\mathcal{N}$ as realizations of $\mathcal{N}$ independent and identically distributed Gaussian random variables $X_n$ and calculating the distribution of their sums, one expects to find a Gaussian distribution with the same mean and variance as the $X_n$'s (according to the Central Limit Theorem). This is indeed what happens for many chaotic dynamical systems studied to date which are \textit{ergodic}, i.e. almost all their orbits (except for a set of measure zero) pass arbitrarily close to any point of the constant energy manifold, after sufficiently long enough times. In these cases, at least one of the Lyapunov exponents \cite{Benettin1980a,Benettin1980b,Eckmann1985,Skokos2010} is positive, stable periodic orbits are absent and the constant energy manifold is covered uniformly by chaotic orbits, for all but a (Lebesgue) measure zero set of initial conditions.

In the galactic models treated in this paper, we focus on weakly chaotic regimes and demonstrate by means of numerical experiments that pdfs of sums of their variables  \textit{do not} rapidly converge to a Gaussian, but are well approximated, for long integration times, by the so-called $q$-Gaussian distribution \cite{Tsallisbook2009}
\begin{equation}
P(s)=a \exp_q({-\beta s^2})\equiv a\biggl[1-(1-q)\beta s^{2}\biggr]^{\frac{1}{1-q}} \nonumber
\end{equation}
where the index $q$ satisfies $1<q<3$, $\beta$ is an arbitrary parameter and $a$ is a normalization constant. At longer times, of course, chaotic orbits eventually ``seep out'' from small regions to larger chaotic seas, where obstruction by islands and cantori is less dominant and the dynamics is more uniformly ergodic. This transition is signaled by the $q$-index of the distribution (\ref{q_gaussian}) decreasing towards $q=1$, which represents the limit at which the pdfs become Gaussian.

In what follows, we begin in Section \ref{Model_gal} by presenting our galaxy barred galaxy model with all its components and parameters. In Section \ref{CLT_approach}, we describe the approach of the $q$-Gaussian distributions and discuss the details of their computation. Sections \ref{2dof_distr}, \ref{3dof_distr} are devoted to the application of these pdfs and the demonstration of their effectiveness in distinguishing between strong and weak chaos in 2 and 3 dof cases of the galactic potential, using also power spectra to support our findings. Finally, we conclude with a summary and discussion of our results in Section \ref{conclusions}.

\section{The barred galaxy model}\label{Model_gal}

Let us consider the motion of test particles in a rotating ``mean field'' model of a barred galaxy potential $V(x,y,z)$ governed by the 3 dof Hamiltonian
\begin{equation}\label{eq:Hamilton}
   H=\frac{1}{2} (p_{x}^{2}+p_{y}^{2}+p_{z}^{2})+ V(x,y,z) -
   \Omega_{b} (xp_{y}-yp_{x})=E_j.
\end{equation}
The bar rotates around its $z$-axis (the short axis), while the $x$-direction rotates along the major axis and the $y$ along the intermediate axis of the bar. $p_{x},\;p_{y}$ and $p_{z}$ are the canonically conjugate momenta, $\Omega_{b}$ represents the pattern speed of the bar and $E_j$ is the total constant energy of the orbit in the rotating frame of reference (namely the Jacobi integral). The corresponding equations of motion are
\begin{flalign}\label{eq_motion}
 \dot{x}& =  p_{x} + \Omega_{b} y,& \quad  \dot{y}& =  p_{y} - \Omega_{b}x,&    \dot{z}& = p_{z},&\\
 \dot{p_{x}}& =  -\frac{\partial V}{\partial x} + \Omega_{b} p_{y},&  \dot{p_{y}}& = -\frac{\partial V}{\partial y} - \Omega_{b} p_{x},&  \dot{p_{z}}& = -\frac{\partial V}{\partial z}.& \nonumber
\end{flalign}
The potential consists of three components $V=V_D+V_S+V_B$:
\begin{enumerate}
 \item The disc, which is represented by a Miyamoto-Nagai disc \cite{Miy:1975}:
 \begin{equation}\label{Miy_disc}
  V_D=-\frac{GM_{D}}{\sqrt{x^{2}+y^{2}+(A+\sqrt{z^{2}+B^{2}})^{2}}},
\end{equation}
where $M_{D}$ is the total mass of the disc, $A$ and $B$ are its horizontal and vertical scale-lengths, and $G$ is the gravitational constant.
\item The bulge, which is modeled by a Plummer sphere \cite{Plum} whose potential is:
\begin{equation}\label{Plum_sphere}
    V_S=-\frac{G M_{S}}{\sqrt{x^{2}+y^{2}+z^{2}+\epsilon_{s}^{2}}}.
\end{equation}
$\epsilon_{s}$ is the scale-length of the bulge and $M_{S}$ is its total mass.
\item The triaxial Ferrers bar \cite{Fer}, the density $\rho(x)$ of which is
\begin{equation}\label{Ferr_bar}
  \rho(x)=\begin{cases}\rho_{c}(1-m^{2})^{2}, & m<1  \\
              \qquad 0, & m\geq1 \end{cases},
\end{equation}
where $\rho_{c}=\frac{105}{32\pi}\frac{G M_{B}}{abc}$ is the central density, $M_{B}$ is the total mass of the bar and
\begin{equation}\label{Ferr_m}
  m^{2}=\frac{x^{2}}{a^{2}}+\frac{y^{2}}{b^{2}}+\frac{z^{2}}{c^{2}},
\qquad a>b>c> 0,
\end{equation}
with $a,b$ and $c$ being the semi-axes. The corresponding potential is:
\begin{equation}\label{Ferr_pot}
    V_{B}= -\pi Gabc \frac{\rho_{c}}{n+1}\int_{\lambda}^{\infty}
    \frac{du}{\Delta (u)} (1-m^{2}(u))^{n+1},
\end{equation}
where
\begin{equation}\label{mu2}
m^{2}(u)=\frac{x^{2}}{a^{2}+u}+\frac{y^{2}}{b^{2}+u}+\frac{z^{2}}{c^{2}+u},
\end{equation}
\begin{equation}\label{Delta}
\Delta^{2} (u)=({a^{2}+u})({b^{2}+u})({c^{2}+u}),
\end{equation}
$n$ is a positive integer (with $n=2$ for our model) and $\lambda$ is the
unique positive solution of
\begin{equation}\label{mu2_lamda}
    m^{2}(\lambda)=1,
\end{equation}
outside of the bar ($m \geq 1$) and $\lambda=0$ inside the bar. The corresponding forces are given analytically in \cite{Pfe:1}.
\end{enumerate}
Throughout the paper, we use the following parameters: $G$=1, $\Omega_{b}$=0.054 (54
$km\cdot sec^{-1} \cdot kpc^{-1}$), $a$=6, $b=$1.5, $c$=0.6, $A$=3, $B$=1,
$\epsilon_{s}$=0.4, $M_{B}$=0.1, $M_{S}$=0.08, $M_{D}$=0.82, both for the 2 and 3 dof versions. The units used are: 1 $kpc$ (length), 1000 $km\cdot
sec^{-1}$ (velocity), 1 $Myr$ (time), $2 \times 10^{11} M_{\bigodot}$ (mass).
The total mass $G(M_{S}+M_{D}+M_{B})$ is set equal to 1 and the corotation radius is equal to $R_c=6.13$.

The above model was used extensively in orbital studies \cite{PSA02,PSA03a,PSA03b,SPA02a,SPA02b}, in which the stability of its main periodic orbits was thoroughly investigated, as several of its parameters were varied. In \cite{Manos_etal:2008,ManAthan:2009,MA11a}, a dynamical study of regular, chaotic and weakly chaotic motion was presented (both in phase and configuration space) for different distributions of initial conditions and values of the parameters of the bar. Moreover, the SALI/GALI method \cite{Sk_sali:2001,SBA_gali:2007} and frequency spectrum analysis were applied to discriminate between strongly and weakly chaotic orbits in \cite{ManosPhD,MA11b}. The latter are of great importance from an astronomical point of view, since they can last for long time scales and behave quite ``regularly'' before revealing their inherent chaotic nature.

In our study, the maximal integration of the orbits is typically set to $t_f=10000\;Myr$ (10 billion years), which corresponds to a time less than (but of the order of) one Hubble time. Nevertheless, we often extend our simulations to longer times to study the dynamics of our solutions in greater detail.

\section{Statistical distributions of chaotic quasi-stationary states and their computation}\label{CLT_approach}

The approach we shall adopt is in the spirit of the Central Limit Theorem: Using the solutions of the equations of motion (\ref{eq_motion}) we construct probability density functions (pdfs) of suitably rescaled sums of an observable function $\eta_i=\eta(t_i)$, which depends linearly on the position coordinates of the solution and is calculated at the ends of $M$ successive time intervals $\Delta t=t_i-t_{i-1},~~~i=1,\ldots,M$. If the motion is chaotic and $\Delta t$ is relatively ``large'', the $\eta(t_i)$ can be assumed to be independent and identically distributed random variables (in the large $M$ limit). Thus, evaluating their sum
\begin{equation}\label{sums_CLT}
S_M^{(j)}=\sum_{i=1}^M\eta_i^{(j)}
\end{equation}
for $j=1,\ldots,N_{ic}$ initial conditions, we can study the statistics of these variables centered about their mean value $\langle S_M^{(j)}\rangle$ and rescaled by their standard deviation $\sigma_M$ as
\begin{equation}\label{s_M_j variables}
s_M^{(j)}\equiv\frac{1}{\sigma_M}\Bigl(S_M^{(j)}-\langle S_M^{(j)}\rangle \Bigr)=\frac{1}{\sigma_M}\Biggl(\sum_{i=1}^M\eta_i^{(j)}-\frac{1}{N_{ic}}\sum_{j=1}^{N_{ic}}\sum_{i=1}^{M}\eta_i^{(j)}\Biggl)
\end{equation}
where
\begin{equation}
\sigma_M^2=\frac{1}{N_{ic}}\sum_{j=1}^{N_{ic}}\Bigl(S_M^{(j)}-\langle S_M^{(j)}\rangle \Bigr)^2=\langle S_M^{(j)2}\rangle -\langle S_M^{(j)}\rangle^2.
\end{equation}
Plotting the normalized histogram of the probabilities $P(s_M^{(j)})$ as a function of $s_M^{(j)}$, we approximate the resulting pdfs by $q$-Gaussian functions of the form
 \begin{equation}\label{q_gaussian}
P(s)=a \exp_q({-\beta s^2})\equiv a\biggl[1-(1-q)\beta s^{2}\biggr]^{\frac{1}{1-q}},
\end{equation}
 where $q$ is the so-called entropic index, $\beta$ is a free parameter and $a$ a normalization constant \cite{Tsallisbook2009}. If the motion is strongly chaotic, the pdfs are expected to converge to the well-known Gaussian, which represents the limit of (\ref{q_gaussian}) as $q\rightarrow 1$. However, in regimes of weak chaos it has been found in many studies that the orbits form quasi-stationary states, which can be very long-lived and whose pdfs are well-fitted by $q$-Gaussian functions with $q$ well above unity \cite{ABB_QSS,Baldovin2004a,Baldovin2004b,Tsallisbook2009}. It can be shown that (\ref{q_gaussian}) is normalized by setting
\begin{equation}\label{beta-$q$-Gaussian}
\beta=a\sqrt{\pi}\frac{\Gamma\Bigl(\frac{3-q}{2(q-1)}\Bigr)}{(q-1)^{\frac{1}{2}}\Gamma\Bigl(\frac{1}{q-1}\Bigr)}
\end{equation}
($\Gamma$ is the Euler Gamma function), implying that the allowed values of $q$ are $1<q<3$. The index $q$ is connected with the Tsallis entropy \cite{Tsallisbook2009}
\begin{equation}\label{Tsallis entropy}
S_q=k\frac{1-\sum_{i=1}^W p_i^q}{q-1}\mbox{ with }\sum_{i=1}^W p_i=1
\end{equation}
where $i=1,\ldots,W$ counts the microstates of the system, each occurring with a probability $p_i$ and $k$ is the so-called Boltzmann universal constant. Just as the Gaussian distribution represents an extremal of the Boltzmann-Gibbs entropy $S_{BG}\equiv S_1=k\sum_{i=1}^W p_i\ln p_i$, so is the $q$-Gaussian (\ref{q_gaussian}) derived by optimizing the Tsallis entropy (\ref{Tsallis entropy}) under appropriate constraints.

Systems characterized by the Tsallis entropy occur very frequently in a variety of applications and obey statistics different than Boltzmann-Gibbs, in the sense that their entropy is nonadditive and generally nonextensive \cite{Tsallisbook2009,TsallisTirnakli2010}. Examples falling in this class are physical systems governed by long range forces, like self-gravitating systems of finitely many mass points, interacting black holes and ferromagnetic spin models, in which correlations decay by power laws rather than exponentially \cite{Tsallisbook2009}.

Let us now describe the numerical approach we follow to calculate the above pdfs. First we select $\eta(t)$ as one (or a linear combination) of the position coordinates $x(t),y(t),z(t)$ of an orbit which we wish to identify as strongly or weakly chaotic.
Choosing then an integration interval $0\leq t\leq t_{f}$ sufficiently long to obtain reliable statistics, we divide $t_f$ into $N_{ic}$ equally spaced, consecutive time windows, which are long enough to contain a significant part of the orbit. Next, we subdivide each such window into a number $M$ of equally spaced subintervals and calculate the sums (\ref{sums_CLT}) at the ends of these subintervals as described above.

In this way, we treat the point at the beginning of every time window as a new initial condition and repeat this process $N_{ic}$ times to obtain as many sums as possible for the time interval $t_f$. Consequently, at the end of the integration, we evaluate the $N_{ic}$ rescaled quantities $s_M^{(j)}$ (\ref{s_M_j variables}) and plot the histogram $P(s_M^{(j)})$ of their distribution. If we are in a regime where these distributions are well-fitted by $q$-Gaussians for fairly long times, the orbit can be classified as weakly chaotic. If, on the other hand, the orbit is strongly chaotic, its pdf is seen to approach a Gaussian, already for relatively short time intervals not exceeding, in our case, the age of the universe.

\section{Statistical distributions of the 2 dof model}\label{2dof_distr}

Let us start by analyzing the 2 dof version of our galactic model (\ref{eq:Hamilton}), setting $z(0)=p_z(0)=0$ initially, whence $z(t)=p_z(t)=0$ for all $t$. In particular, we are interested in identifying the dynamical behavior of the following two orbits:
\begin{itemize}
\item[(i)] The 2-Dimensional Strongly Chaotic (2DSC) orbit, with initial condition\\ $(x,y,p_x,p_y)=(0,-0.625,-0.314512,-0.24)$ and $E_j=-0.3$, \\ and
\item[(ii)] The 2-Dimensional Weakly Chaotic (2DWC) orbit, with initial condition\\ $(x,y,p_x,p_y)=(0,-0.625,-0.002,-0.24)$ and $E_j=-0.36$.
\end{itemize}
These orbits for the 2 dof case (and the next two of section 5 for the 3 dof case) are selected from a vast number of trajectories (see [24, 27]), as representative examples of weakly and strongly chaotic motion to which we can apply our pdf approach and link the results with questions of interest to dynamical astronomy. The authors in [24, 27] used dynamical chaos detectors on thousands of orbits, in order to investigate and chart the fractions of chaotic and regular regimes of the phase space and then link these fractions to the main parameters of the bar component. Since it was evident from these studies that there is a significant amount of weakly chaotic orbits that appear ``regular'' for astronomically bounded time scales, we decided to revisit this question in the present paper from a statistical point of view.
\begin{figure}[!ht]
\centering
\includegraphics[scale=0.9]{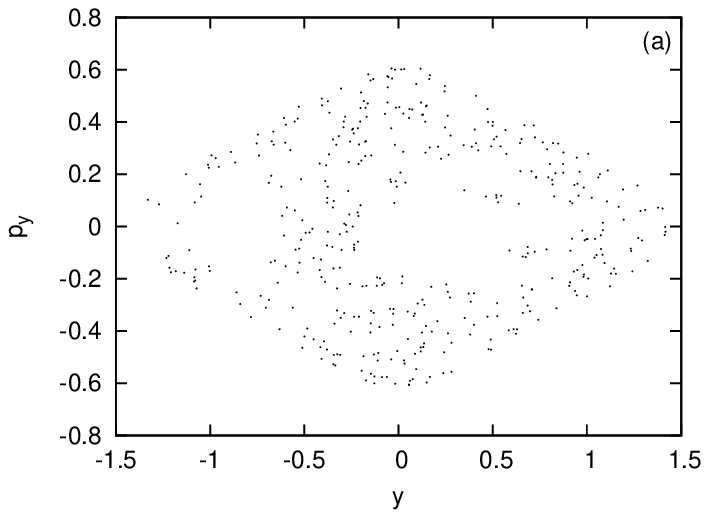}
\includegraphics[scale=0.9]{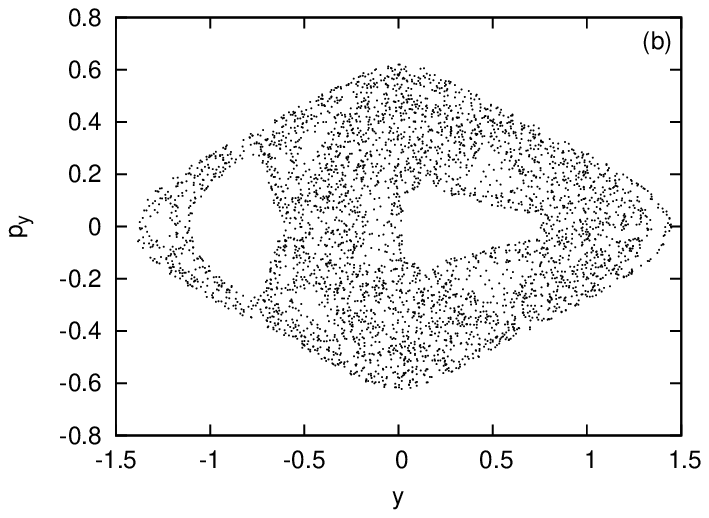}\\
\includegraphics[scale=0.9]{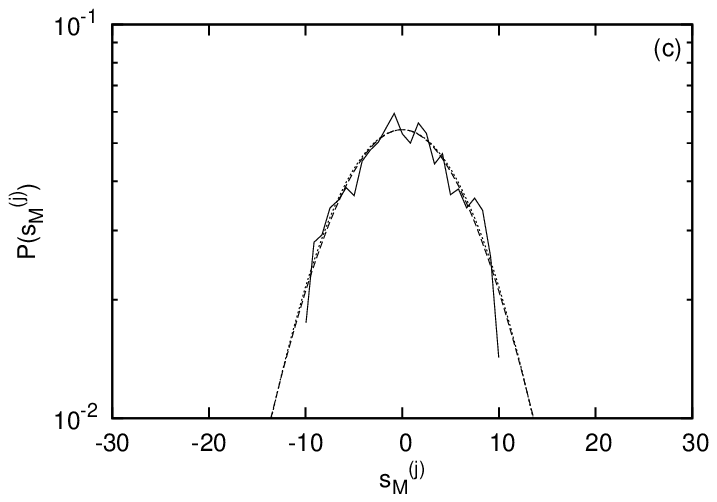}
\includegraphics[scale=0.9]{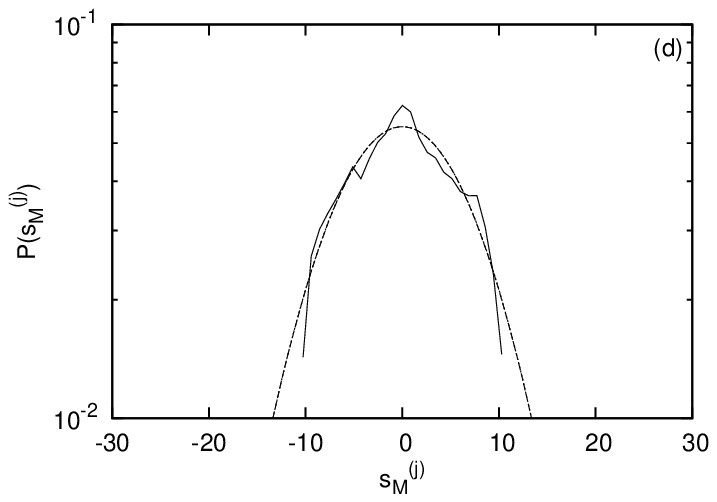}
\caption{Both rows correspond to $t_f=10000$ (left panels) and
$t_f=100000$ (right panels). (a) and (b): surface of section plots $(y,p_y)$ with $p_x>0,\,x=0$ for the 2DSC orbit of the 2 dof Hamiltonian system (\ref{eq:Hamilton}), showing evidence of widespread chaos (the horizontal scale is in $kpc$ in both top panels). (c) and (d): Linear-log scale plots of the pdf of the numerically computed orbit (solid curve) and a Gaussian distribution (dashed curve) fitted to the data for the observable $\eta=x+y$. In (c) we use $N_{ic}=4000$ time windows and $M=50$ terms in the computation of the sums and the numerical fitting with the $q$-Gaussian (\ref{q_gaussian}) gives $q\approx1.095$ with $\chi^2\approx0.0003$. (d) corresponds to $N_{ic}=20000$ and $M=10$ and gives $q\approx0.97$ with $\chi^2\approx0.00033$. Since in both panels (c) and (d) the pdfs are well approximated by Gaussians, we conclude that 2DC is a strongly chaotic orbit} \label{fig1}
\end{figure}

\begin{figure}[!ht]
\centering
\includegraphics[scale=0.9]{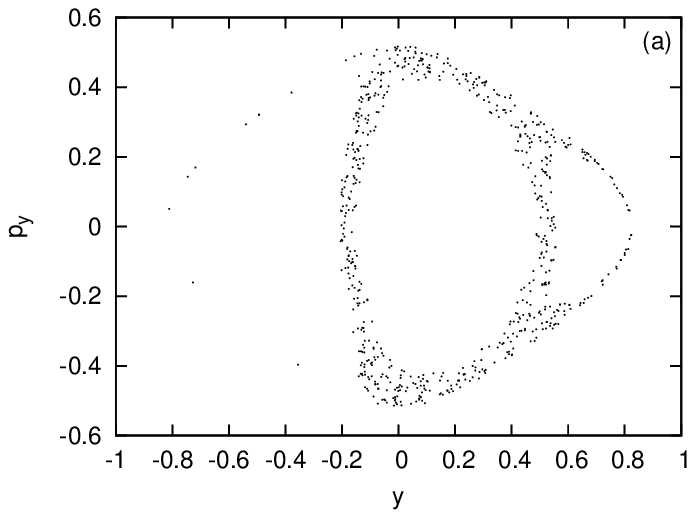}\hspace{0.cm}
\includegraphics[scale=0.9]{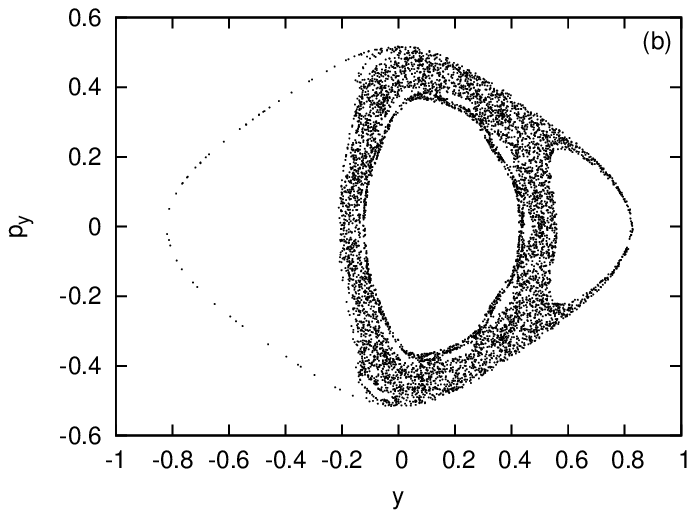}\\
\includegraphics[scale=0.9]{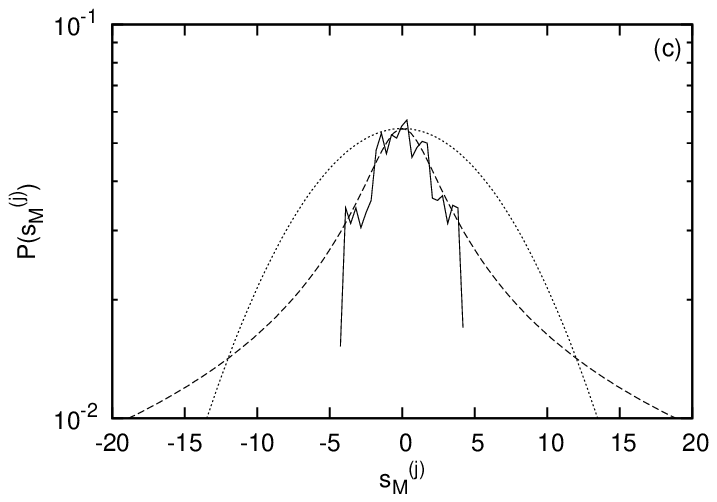}\hspace{0.cm}
\includegraphics[scale=0.9]{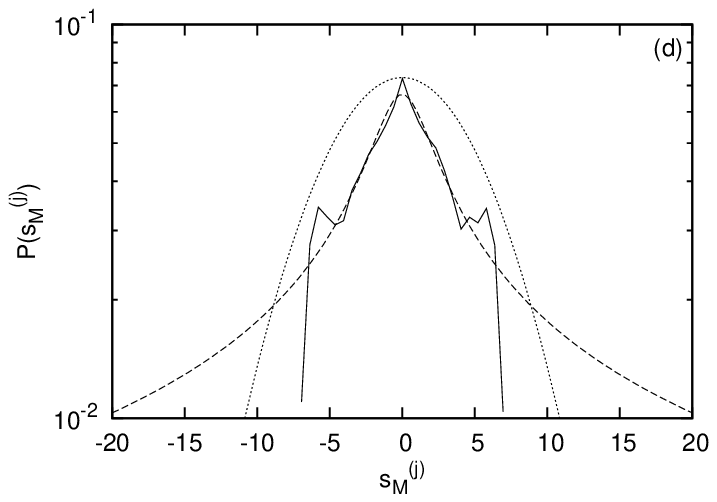}
\caption{We use again final integration times $t_f=10000$ for the left panels and $t_f=100000$ for the right ones. (a) and (b): surface of section plots of  $(y,p_y)$ with $p_x>0,\,x=0$ for the orbit 2DWC of the 2 dof galactic system (\ref{eq:Hamilton}) (the horizontal scale is in $kpc$ in both top panels). (c) and (d): Linear-log scale plots of numerical (solid curve), $q$-Gaussian (dotted curve) and Gaussian distributions (dashed curve) for the observable $\eta=x+y$, showing evidence of weak chaos. In (c) we have used $N_{ic}=4000$ time windows and $M=50$ terms in the computation of the sums and the numerical fitting with the $q$-Gaussian (\ref{q_gaussian}) gives $q\approx3.52$ with $\chi^2\approx0.00074$. In (d) we have taken $N_{ic}=25000$ and $M=50$. Here, the numerical fitting gives $q\approx3.539$ with $\chi^2\approx0.00057$. Of course, the fact that the entropic parameter is $q>3$ implies that these distribution functions are not normalizable and hence cannot represent $q$-Gaussians for this orbit.} \label{fig2}
\end{figure}

\begin{figure}[!ht]
\centering
\includegraphics[width=8cm]{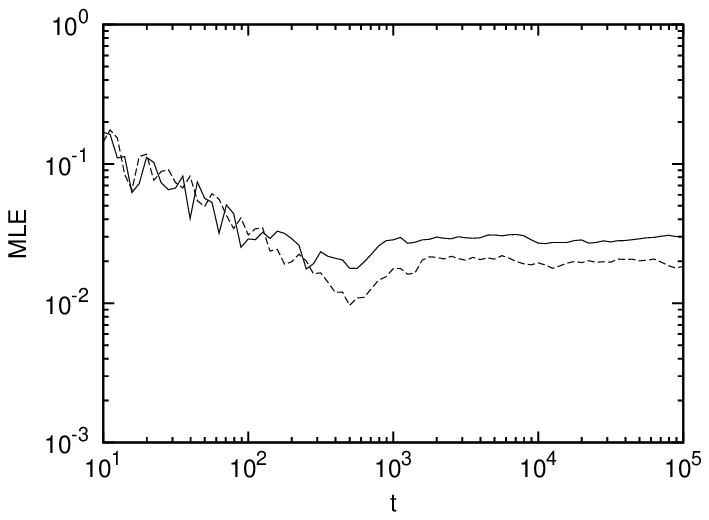}
\caption{Time evolution of the MLE of the strongly chaotic orbit 2DSC showing a tendency to converge to $\sigma \simeq 0.03011$ (solid curve) and the MLE of the weakly chaotic 2DWC orbit approaching the value $\sigma \simeq 0.01489$ (dashed curve), for the 2 dof Hamiltonian system (\ref{eq:Hamilton}).} \label{fig3}
\end{figure}

In Fig.~\ref{fig1} the left panels correspond to the integration time $t_f=10000$ and the right panels to $t_f=100000$. Figs.~\ref{fig1} (a) and (b) represent surface of section plots $(y,p_y)$ with $p_x>0,\,x=0$ for the orbit 2DSC, with $E_j=-0.3$, showing clear signs of widespread chaos as the orbit ``fills out'' rather uniformly a wide chaotic domain. Figs.~\ref{fig1} (c) and (d), on the other hand, depict the numerically computed pdfs (solid curve) for the observable $\eta=x+y$ and the Gaussian distributions (dashed curve) that best fit the data.  As is evident from these results the proximity of the computed pdfs to a Gaussian, already at $t_f=10000$, suggests that the 2DSC orbit can be classified as strongly chaotic.

Let us now turn to our analysis of orbit 2DWC in Fig.~\ref{fig2}, with $E_j=-0.36$ using again integration times $t_f=10000$ for the left panels and $t_f=100000$ for the right panels. Note on the surface of section panels of Fig.~\ref{fig2}(a) and (b) that the 2DWC orbit fills a region that is considerably more limited than that of the 2DSC orbit (see Figs.~\ref{fig1}(a),(b)), containing several islands to which the orbit apparently sticks for long times. Indeed, when we compute the sum pdfs of the variable $\eta=x+y$ for this orbit in Figs.~\ref{fig2}(c) and(d), we find that they are well-approximated by $q$-Gaussians that are quite far from a Gaussian, already for $t_f=10000$, thus suggesting that 2DWC cannot be regarded as a strongly chaotic orbit. Note, however, that ``stickiness'' is just one aspect of this form of chaos, since the 2DWC orbit spends only part of its evolution in the vicinity of a number of major ``islands''.

It is important to note, however, that the distribution functions plotted in Figs.~\ref{fig2}(c) and (d) have $q>3$ and thus are not normalizable and cannot be used to represent $q$-Gaussians for this orbit. They only serve to show that the dynamics is certainly far from Gaussian and hence the 2DWC orbit cannot be classified as strongly chaotic.

In Fig.~\ref{fig3} we plot the time evolution of the MLE of the two orbits of the 2 dof system. We find that the MLE of the strongly chaotic chaotic orbit 2DSC (solid curve) converges to the value $\sigma \simeq 0.03011$, while the MLE of the weakly chaotic orbit 2DWC (dashed curve) tends to $\sigma \simeq 0.01489$. Although this value is quite smaller than the one for the 2DSC orbit, these results demonstrate that merely knowing the values of the MLEs is \emph{not} sufficient for determining whether an orbit belongs to the weakly or strongly chaotic family.

\section{Statistical distributions of the 3 dof system} \label{3dof_distr}

Let us now focus on the 3 dof  system (\ref{eq:Hamilton}) and carry out a similar study as in the previous section for the orbits:
\begin{itemize}
\item[(iii)] The 3-Dimensional Strongly Chaotic (3DSC) orbit, with initial condition\\ $(x,y,z,p_x,p_y,p_z)=(0.5875,0,1.291670,0,0,0)$ with $E_j=-0.2792149022676664$, \\and
\item[(iv)] The 3-Dimensional Weakly Chaotic (3DWC) orbit, with initial condition\\ $(x,y,z,p_x,p_y,p_z)=(2.35,0,0.08883,0,0.133330,0)$ and $E_j=-0.2852654501087482$
\end{itemize}
of the 3 dof Hamiltonian system (\ref{eq:Hamilton}).

\begin{figure}[!ht]
\centering
\includegraphics[scale=0.9]{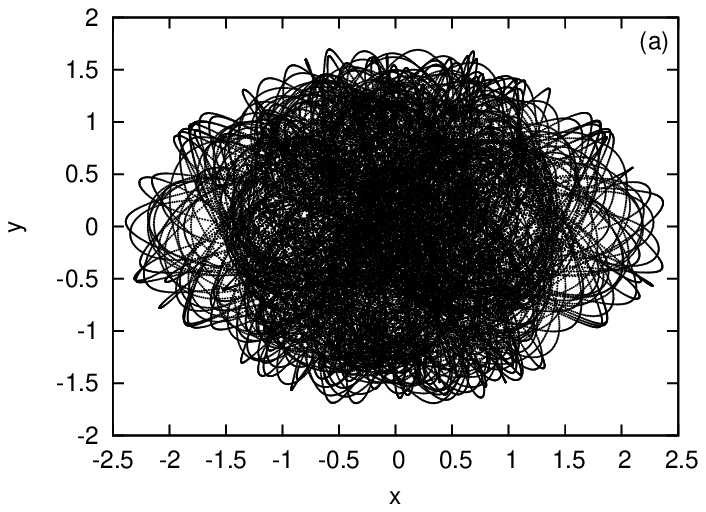}
\includegraphics[scale=0.9]{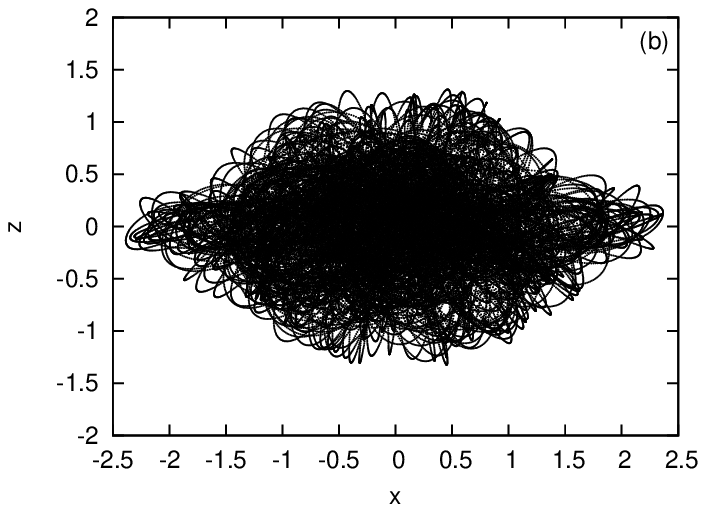}\\
\includegraphics[scale=0.9]{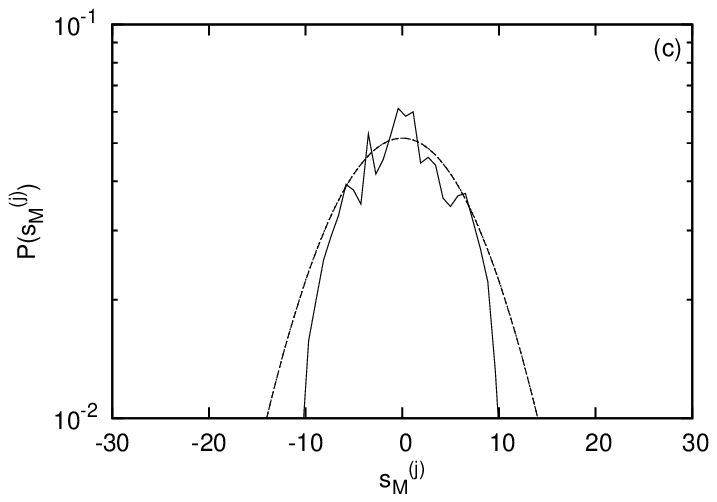}
\includegraphics[scale=0.9]{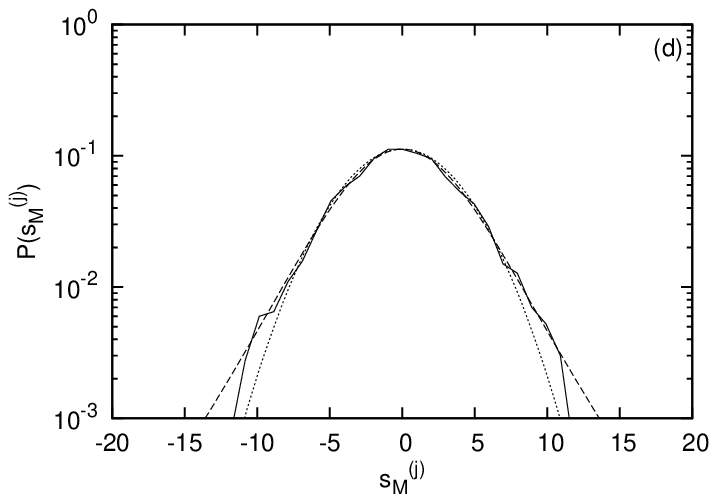}\\
\includegraphics[scale=0.9]{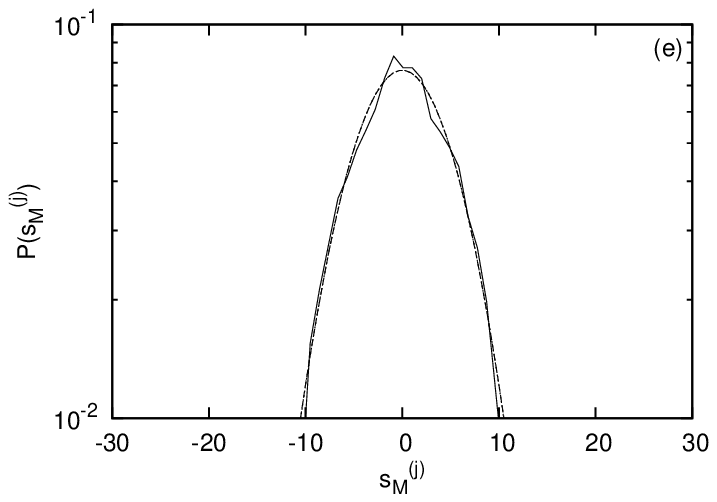}
\includegraphics[scale=0.9]{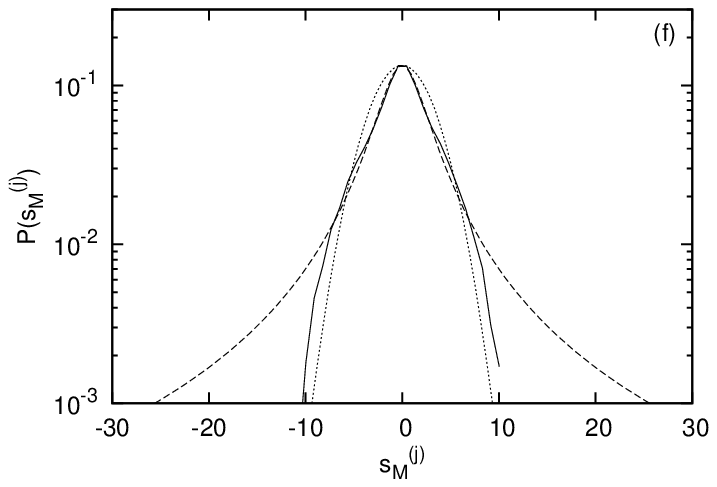}
\caption{Projections of the orbit 3DSC of the 3 dof Hamiltonian system for integration time $t_f=10000$ in (a) the $(x,y)$ plane and (b) the $(x,z)$ plane (the axes' scales are in $kpc$ in both top panels). Second row: Plots in linear-log scale of numerical (solid curve), $q$-Gaussian (dotted curve) and Gaussian (dashed curve) distributions of the same orbit, (c) for the variable $\eta=x+y$, with $N_{ic}=4000$, $M=50$ and (d) $\eta=z$, for $N_{ic}=4000$ and $M=50$, both for $t_f=10000$. Here, the numerical fitting with (\ref{q_gaussian}) gives $q\approx1.25$ with $\chi^2\approx0.00017$. In the third row we have used $t_f=100000$ and plotted pdfs for: (e) the variable $\eta=x+y$, $N_{ic}=10000$ and $M=100$, fitted by a $q$-Gaussian with $q\approx0.95$ and $\chi^2\approx0.00029$, and (f) the $\eta=z$ variable, with $N_{ic}=10000$, $M=100$, where the numerical fitting parameters are $q\approx1.94$ and $\chi^2\approx0.00035$.} \label{fig4}
\end{figure}

Figs.~\ref{fig4}(a) and (b) show projections of the orbit 3DSC in the $(x,y)$ and $(x,z)$ planes respectively for the integration time interval $t_f=10000$. Evidently, no sign of any regularity is observed here and the plots indicate that the orbit might be strongly chaotic. Indeed, as we see in Figs.~\ref{fig4}(c) and (d), both for $t_f=10000$, pdf plots of the variables $\eta=x+y$ in (c) and $\eta=z$ in (d) provide evidence of strong chaos, as both distributions are well-approximated by Gaussians, whose $q$ is close to unity. Panels (e) and (f) show pdf plots for the variables $\eta=x+y$ and $z$ respectively but now for integration time $t_f=100000$. Here also, the numerically fitted curves are close to Gaussians.

Note, however, that the upper part of the numerical data in Fig.~\ref{fig4}(f), is also approximated rather well by a $q$-Gaussian with $q\approx1.94$, which indicates a peculiar property of the $z$ projection of the dynamics, which will be discussed in detail below. Still, based on these findings, we suggest that the 3DSC orbit be classified as strongly chaotic.

\begin{figure}[!ht]
\centering
\includegraphics[scale=0.9]{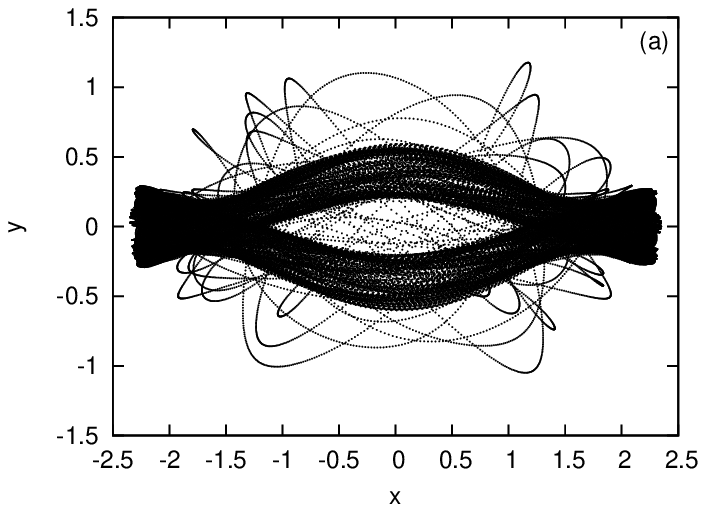}
\includegraphics[scale=0.9]{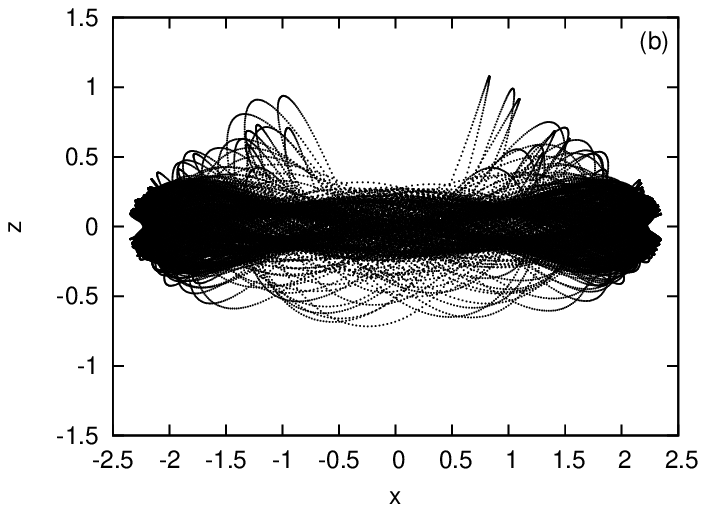}\\
\includegraphics[scale=0.9]{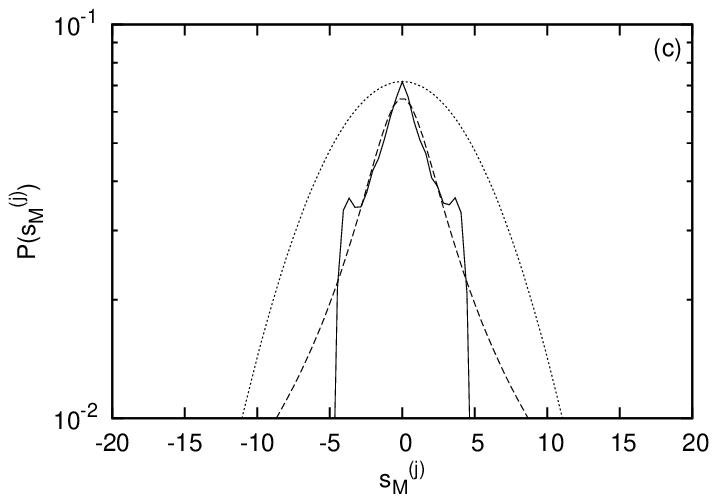}
\includegraphics[scale=0.9]{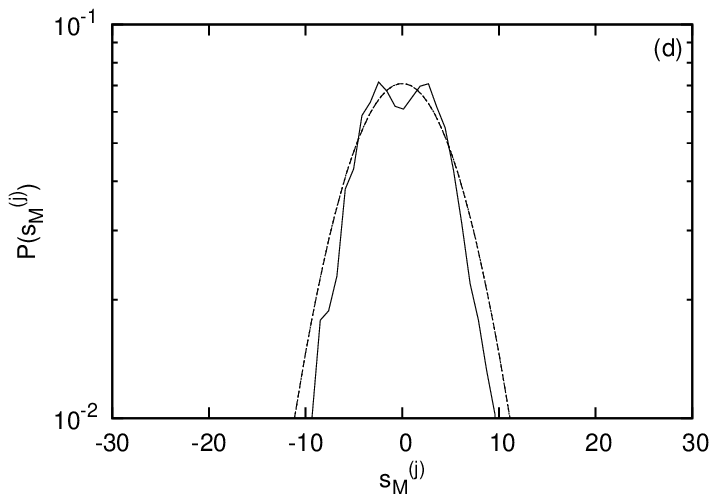}\\
\includegraphics[scale=0.9]{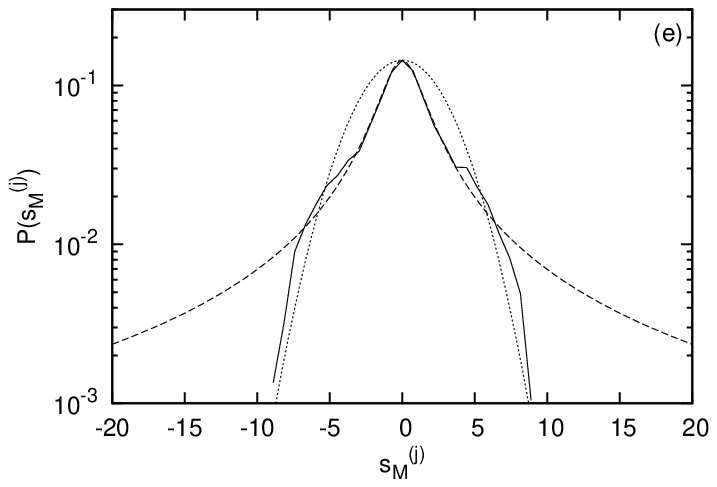}
\caption{First row: Projections of the orbit 3DWC of the 3 dof Hamiltonian system (\ref{eq:Hamilton}) for final integration time $t_f=10000$ (a) on the $(x,y)$ plane and (b) on the $(x,z)$ plane (the axes' scales are in $kpc$ in both top panels). Second row: Plot in linear-log scales of numerical (solid curve), $q$-Gaussian (dotted curve) and Gaussian distributions (dashed curve) for $t_f=100000$. (c) corresponds to  $\eta=x+y$, $N_{ic}=20000$, $M=100$ and presents numerical fittings with a $q$-Gaussian (\ref{q_gaussian}), yielding $q\approx2.464$ with $\chi^2\approx0.00101$. In case (d), however, which corresponds to $\eta=z$, $N_{ic}=4000$ and $M=50$ the data are insufficient to provide reliable estimates. Studying the variable $\eta=z$ for longer times $t_f=100000$, with $N_{ic}=20000$ and $M=100$, a satisfactory numerical fitting by a $q$-Gaussian is obtained giving $q\approx2.262$ with $\chi^2\approx0.00033$.} \label{fig5}
\end{figure}

\begin{figure}[!ht]
\centering
\includegraphics[width=8cm]{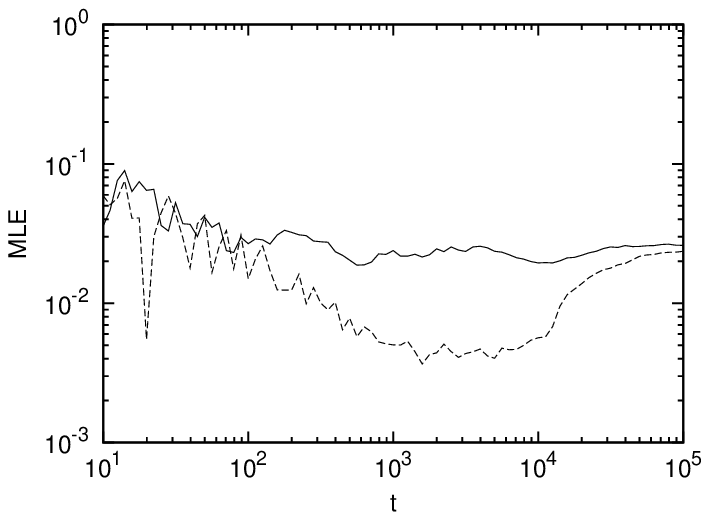}
\caption{The time evolution of the MLE of the chaotic orbit 3DSC ($\sigma \simeq 0.02600$: solid black curve) and of the weakly chaotic orbit 3DWC of the 3 dof Hamiltonian system (\ref{eq:Hamilton}) ($\sigma \simeq 0.0236$: dashed black curve).} \label{fig6}
\end{figure}

Performing now a similar study for the 3DWC solution, we present in Figs.~\ref{fig5}(a),(b) two projections of the orbit for $t_f=10000$, on the $(x,y)$ and $(x,z)$ planes respectively. This orbit is very close to the so-called ``x1'' family of periodic orbits, which have a barred morphological shape and are generally unstable (see \cite{SPA02a}, Table 2 for more details). A certain regularity does appear here, especially in the $(x,y)$ plane of panel (a). Both panels suggest that the motion predominantly occurs in that plane, while frequent chaotic ``excursions'' are also visible in the $z$ direction. In Figs.~\ref{fig5}(c),(d) we plot the numerical (solid curve), $q$-Gaussian (dotted curve) and Gaussian (dashed curve) pdfs of the 3DWC orbit. Fig.~\ref{fig5}(c) corresponds to $\eta=x+y$, for integration time $t_f=100000$, which still indicates a good approximation by a $q$-Gaussian with $q\approx2.464$, suggesting that 3DWC is weakly chaotic.

Interestingly enough, the same cannot be said for $\eta=z$ in Fig.~\ref{fig5}(d), where the data at $t_f=10000$ ``look like'' a Gaussian but are not statistically reliable to draw any conclusion. To investigate the $z$ dynamics further, therefore, we plot in Fig.~\ref{fig5}(e) its pdf for longer integration times ($t_f=100000$) and discover that its central part is well-fitted by a $q$-Gaussian, with $q\approx2.262$ with $\chi^2\approx0.000325$, suggesting that some type of ``order'' is now present. Note also that, in comparison with the 3DSC orbit (see Fig. \ref{fig4}(f)), the statistical distribution of the $z$-coordinate has a higher $q$ value and thus we conclude from all available evidence that 3DWC is a weakly chaotic orbit.

Finally, in Fig.~\ref{fig6}, we present the time evolution of the MLE of the strongly chaotic orbit 3DSC, which does appear to converge rather rapidly to the value $\sigma \simeq 0.02600$ (solid curve). In the case of the weakly chaotic orbit 3DWC, we observe a much slower convergence as the MLE ``dives'' to small values $\approx 3\times10^{-3}$ (dashed curve) at about $t_f\approx 5000$, indicating a weaker form of chaos over this interval and only rises to higher values corresponding to strong chaos for longer times. In fact, in the time interval $0<t<3000$, the 3DWC orbit does display some quasiperiodic features (see also Figs~\ref{fig2}(a),(b)), which will be discussed below, when we study amplitude spectra of the 3DWC orbit.

\begin{figure}[!ht]
\centering
\includegraphics[scale=0.9]{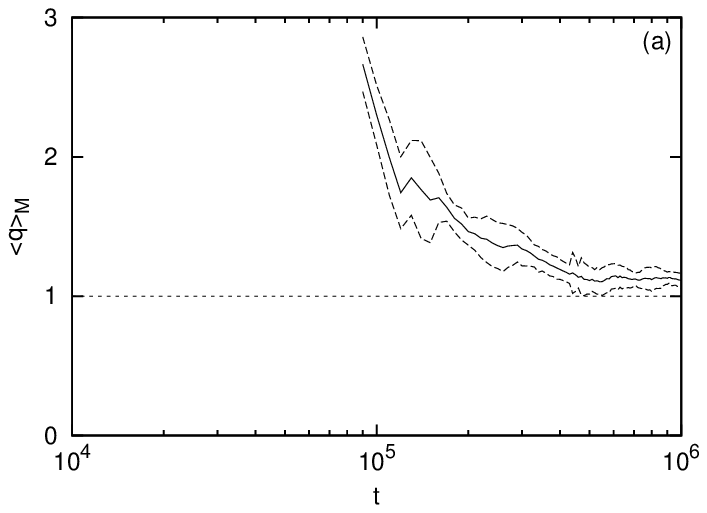}
\includegraphics[scale=0.9]{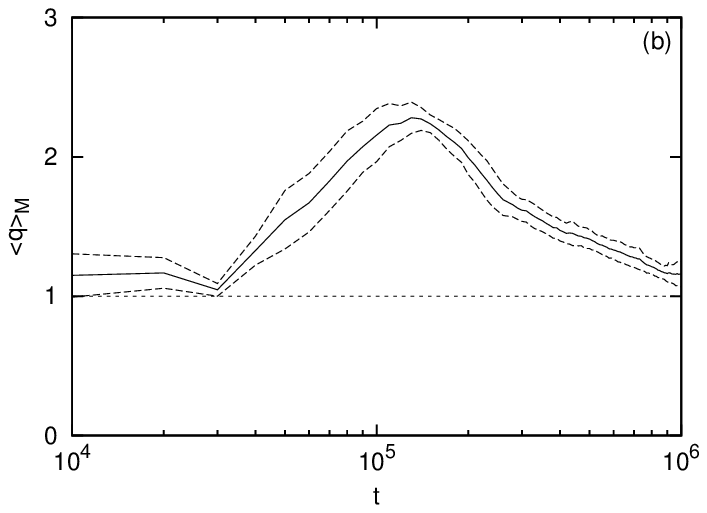}
\caption{The time evolution of the averaged entropic parameter $\langle q\rangle_M$ (see text) for the weakly chaotic orbit 3DWC of the 3 dof system (\ref{eq:Hamilton}) for (a) $\eta=x+y$, $N_{ic}=20000$ and $M=50,100,\ldots,450,500$. (b): Same as in panel (a) for $\eta=z$. The dashed lines correspond to one standard deviation from the average entropic parameter. Note that the horizontal axes are in logarithmic scale and the vertical in linear scale.} \label{fig7}
\end{figure}

Intrigued by the above results, we decided to investigate separately the projections of the 3DWC orbit on the $(x,y)$ plane and the $z$ axis, for longer integration times. In particular, in Fig.~\ref{fig7} we present the time dependence of the entropic index $q$ averaged over $M$ intervals, $\langle q\rangle_M$, taking $M=50,100,\ldots,450,500$ and $N_{ic}=20000$ time windows, computed over intervals $[0,\;T]$ where $T=10000,\;20000,\ldots,\;990000,\;1000000$. The mean and standard deviation of $\langle q\rangle_M$ was found for all pairs of $(N_{ic},M)$ to yield normalized pdfs, i.e. $1\leq q \leq3$ for each time window $[0,\;T]$.

Thus, in Fig.~\ref{fig7}(a) we have plotted $\langle q\rangle_M$ for the observable $\eta=x+y$ (see Fig.~\ref{fig5}(a)) and in Fig.~\ref{fig7}(b) for $\eta=z$ (see Fig.~\ref{fig5}(b)). For $\eta=x+y$, there is quite a long transient time (nearly $85000$ units), for which the motion on the $(x,y)$ projection gives no statistical data (i.e. no pair of $(N_{ic},M)$ that yields $1\leq q \leq3$), since the motion displays strong quasiperiodic features in that time interval. This is also evident by the fact that the MLE of  the orbit (dashed curve in  Fig.~\ref{fig6}) decreases on the average to fairly low values over this interval. A statistically reliable fit is possible only after about $t\approx90000$, yielding $\langle q\rangle_M$ values well above 2 with a trend to approach $q=1$ at longer times (see Fig.~\ref{fig7}). This is in compatible with our claim that 3DWC is a weakly chaotic orbit, for short times.

In the case of the $\eta=z$ observable however, the \emph{opposite} happens: As we see in Fig.~\ref{fig7}(b), in the beginning, $\langle q\rangle_M$ shows a tendency to diminish towards $q=1$ at about $t=30000$, although the statistics is unsatisfactory, since most of the $(N_{ic},M)$ pairs yield $q<1$ or $q>3$. Soon after, however, it increases significantly until it reaches a maximum $q\approx2.28$ at $t=130000$ before it starts to decrease again to $q\approx1.16$ at $t=1000000$. These results are in agreement with our earlier observation that in the $z$ direction the dynamics is initially more strongly chaotic than in the $x,y$ projections, but does show a clear tendency towards weak chaos at later times.

\begin{figure}[!ht]
\centering
\includegraphics[width=3.6cm,height=5cm]{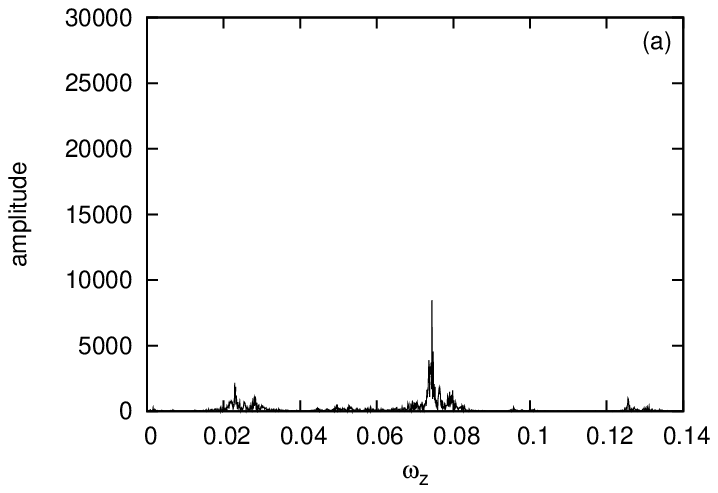}
\includegraphics[width=3.6cm,height=5cm]{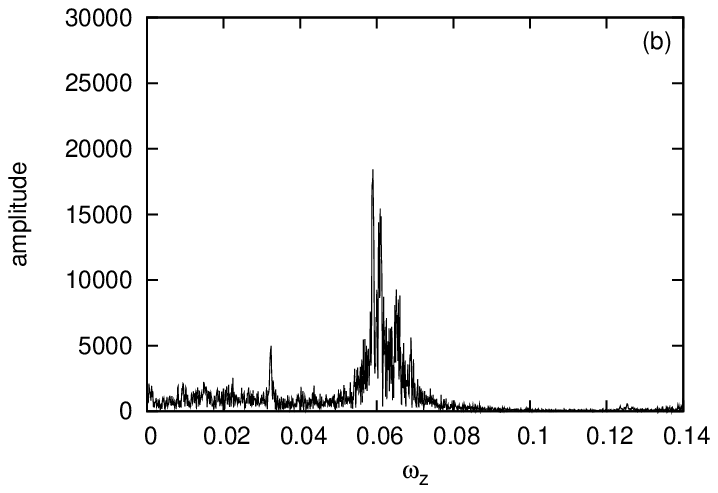}
\includegraphics[width=3.6cm,height=5cm]{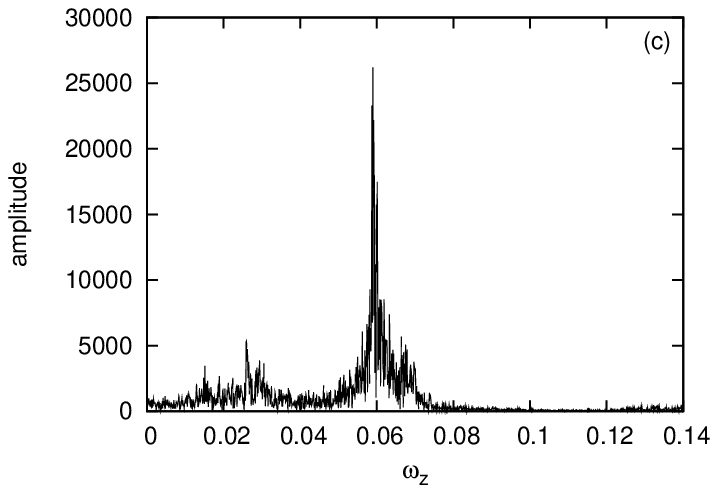}
\includegraphics[width=3.6cm,height=5cm]{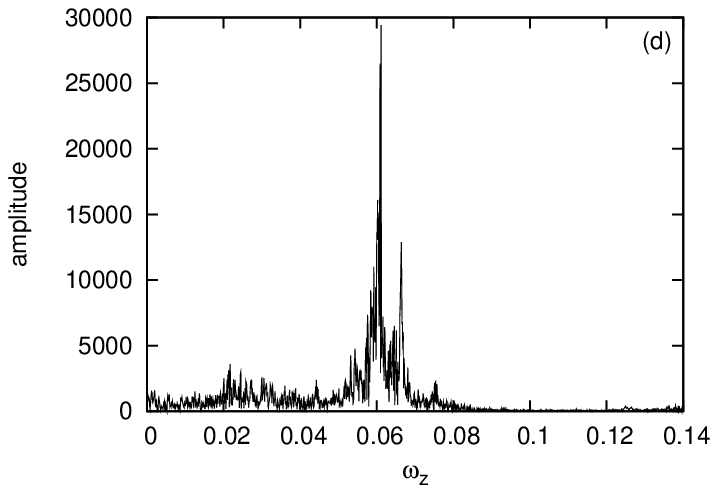}
\caption{Amplitude spectrum of the coordinate component $z$ for the weakly chaotic orbit 3DWC of the 3 dof system, for 4 time windows of length $10000$. (a) Shown here is the spectrum for the time window $[0,\;10000]$ where the amplitude of the main frequency $\omega_z\approx0.07$ is approximately $8500$. (b) Same as in (a) for the time window $[40000,\;50000]$ where the power of the main frequency has now increased to a value $\sim 19000$, a trend that agrees with the increase of $\langle q\rangle_M$ over this time interval. (c) Spectrum is plotted for the time window $[90000,\;100000]$, where the amplitude of the main frequency has further increased to a value of about $27000$, while $\langle q\rangle_M$ increases even more to a value of about $2$. Finally, in panel (d) we plot the same spectrum for $t\in[990000,\;1000000]$.} \label{fig8}
\end{figure}

\begin{figure}[!ht]
\centering
\includegraphics[width=3.6cm,height=5cm]{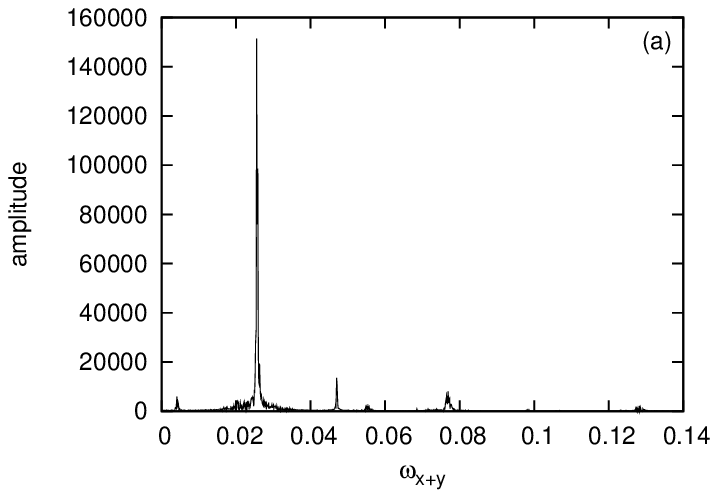}
\includegraphics[width=3.6cm,height=5cm]{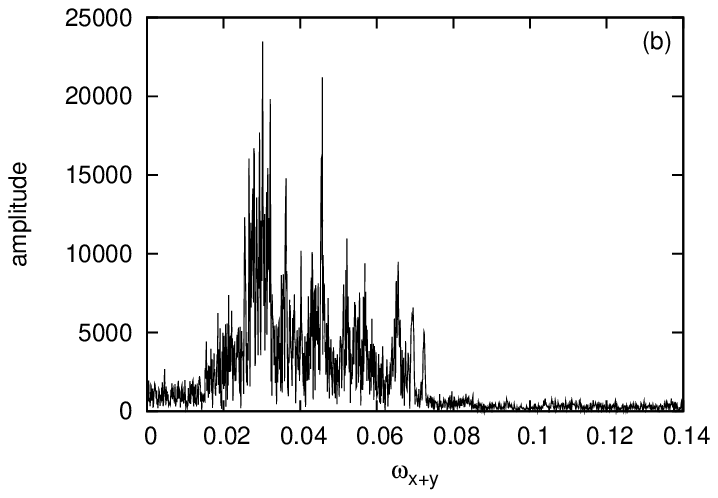}
\includegraphics[width=3.6cm,height=5cm]{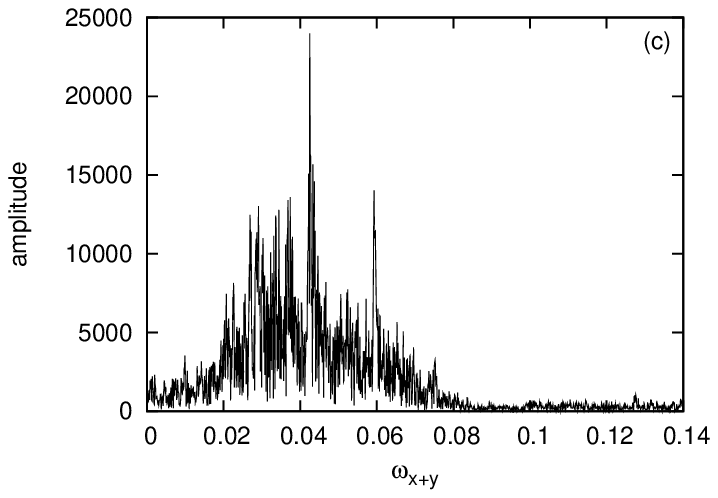}
\includegraphics[width=3.6cm,height=5cm]{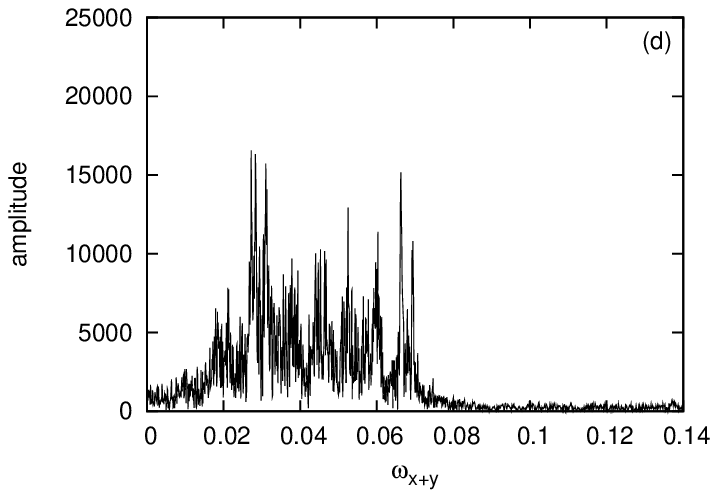}
\caption{Amplitude spectrum of the coordinate component $x+y$ for the weakly chaotic orbit 3DWC of the 3 dof system, for 4 time windows of length $\Delta t=10000$. (a) The spectrum is shown for the time window $[0,\;10000]$ where the amplitude of the main frequency $\omega_{x+y}\approx0.02$ is approximately $150000$.(b) Same as (a) for the time window $[40000,\;50000]$, where the dominant frequency is now almost absent and the total spectrum has spread to a denser set of frequencies. In panel (c), we plot the spectrum for the time window $[90000,\;100000]$. It is in this interval, that $\langle q\rangle_M$ starts gradually to decrease to $1$, the limit at which the distribution becomes Gaussian. Finally, in panel (d) we plot the spectrum for $t\in[990000,\;1000000]$.} \label{fig9}
\end{figure}

This peculiar behavior of the $z$ dynamics is better understood when we compute amplitude spectrum of the $z$ component of the motion, using Fast Fourier Transform (FFT) techniques \cite{NuRe}. Thus, in Fig.~\ref{fig8} we present such spectra for 4 time windows as follows: In Fig.~\ref{fig8}(a) the spectrum is shown for the initial time window $[0,\;10000]$, where the amplitude of the main frequency $\omega_z\approx0.07$ is approximately $8,500$. In Fig.~\ref{fig8}(b), the same plot is presented for the time window $[40000,\;50000]$, with the amplitude of the main frequency increasing to a value $\approx 19000$ (in accordance with the increase of the $\langle q\rangle_M$ parameter found for this interval).

In Fig.~\ref{fig8}(c), the above spectrum is plotted for the time window $[90000,\;100000]$, where the amplitude of the main frequency now increases to about $27000$ (in this time window $\langle q\rangle_M$ has increased even more, reaching a value about $2.1$). Finally, in Fig.~\ref{fig8}(d) we plot the spectrum for $t\in[990000,\;1000000]$ and see that as time increases, the energy of the spectrum is restricted within a very limited set of frequencies around a central $\omega_z$ value, which seems to continuously gain more energy and play a dominant role that is most likely responsible for the weakly chaotic behavior of the $z$ component of the orbit.

To provide further support for our claim that 3DWC is weakly chaotic, we performed the same kind of analysis for its $x+y$ component in Fig.~\ref{fig9}. Fig.~\ref{fig9}(a) shows the frequency spectrum for the time window $[0,\;10000]$, where the amplitude of the main frequency $\omega_{x+y}\approx0.025$ is nearly $150000$. In Fig.~\ref{fig9}(b), the spectrum is presented for the intermediate time window $[40000,\;50000]$, over which the data has spread to a \emph{wider} set of frequencies. In Fig.~\ref{fig9}(c), the spectrum for the time window $[90000,\;100000]$ look similar to panel (b). Interestingly, over this time interval, $\langle q\rangle_M$ still increases, reaching its maximum value $q\approx2.28$ at $t=130000$.

Fig.~\ref{fig9}(d), on the other hand, for $t\in[990000,\;1000000]$ clearly shows that a broad set of low amplitude frequencies still carries the energy of the orbit, without any dominant frequency present, of the type found in Figs.~\ref{fig9}(a)-(c). This further demonstrates that the orbit becomes more strongly chaotic only after very long times, as shown also by the approach of the $\langle q\rangle_M$ index to $q=1$ in Fig.~\ref{fig7}(a).

\section{Summary and conclusions}\label{conclusions}

We have studied statistical distributions of sums of position coordinates for some typical examples of chaotic orbits of a 2 and 3 dof barred galaxy model, aiming to classify them as strongly or weakly chaotic over relatively short time intervals (of the order of one Hubble time). This classification is important, when one wishes to quantify order and chaos in galaxy models, and is not easy to achieve for short times using only local dynamic indicators such as Lyapunov exponents.

More specifically, we have analyzed the statistics of two orbits of a 2 dof barred galaxy potential, having $z(t)=0$ for all time, and two orbits of the 3dof model, whose $z(t)$ components evolve away from zero. If the pdfs produced by the orbits over these intervals are well-approximated by $q$-Gaussian functions, with $q$ significantly larger than 1 ($1<q<3$), we identify the orbits as weakly chaotic. If, on the other hand, $q$ is close to 1, the pdf is well-fitted by a Gaussian and the orbits are said to exhibit strong chaos.

In summary, we have arrived at the following conclusions:

\begin{enumerate}

\item Weak chaos is a concept that encompasses more that the well-known ``stickiness'' phenomenon observed near the boundaries of major ``islands'' of ordered motion or along invariant manifolds of unstable periodic orbits.

\item Statistical methods based on $q$-Gaussian ($1<q<3$) distribution functions can reliably distinguish between strongly and weakly chaotic dynamics in barred galaxy models, over relatively short time intervals.

\item In general, weakly chaotic orbits evolve into wider chaotic domains with Gaussian pdfs corresponding to strong chaos, but over longer time scales exceeding one Hubble time. This shows that weak chaos is an important feature of stellar dynamics that may possibly influence the formation and support of certain galaxy structures.

\item Statistical distributions of different observables do not always show the same dynamical features in different projections. For example, the pdfs of the $\eta=x+y$ and $\eta=z$ variables of the galaxy model studied here, revealed a weaker form of chaotic motion (and for longer time intervals) in the $x,y$ plane compared to the $z$ direction.

\item The averaged entropic index $\langle q\rangle_M$ is a useful parameter for monitoring the nature of the dynamics of different observables. In the case of our 3 dof weakly chaotic orbit for example, the surprisingly distinct time evolution  of $\langle q\rangle_M$ for the $\eta=x+y$ observable, compared with $\eta=z$, reflects the sharply different dynamics of the orbit in the $x,y$ and $z$ projections.

\end{enumerate}

The properties of weak and strong chaos have been observed in many other orbits beyond the four examples treated in the paper. After identifying the properties of the $q$-entropic parameter in typical cases of ``weakly" and ``strongly" chaotic motions, future studies should be made to extend this approach to bigger samples of orbits to provide a more quantitative perspective on their distribution in phase and real space As explained in \cite{Tsallisbook2009}, the occurrence of $q$-Gaussian pdfs allows one to define a new form of partition function and thus develop a different type of (nonextensive) thermodynamics to describe dynamical systems in these regimes.

We believe that the above conclusions and results are not limited to barred galaxy models. It would be interesting, therefore, to apply the statistical methods described in this paper to different galaxy structures to quantify their chaotic regimes in more detail and obtain a more complete picture of their dynamics.

\begin{acknowledgements}
This work was partially supported by a grant from GSRT, Greek Ministry of Education, for the project ``Complex Matter'', awarded under the auspices of the ERA Complexity Network. The authors gratefully thank Prof. L. Drossos and the HPCS Lab of the Technological Educational Institute of Messolonghi, for helping us use their computer facilities to run some of the numerical simulations presented in this paper.
\end{acknowledgements}



\begin{thebibliography}{}

\bibitem{ABB_QSS} Antonopoulos, Ch., Bountis, T., Basios, V.: Quasi-stationary chaotic states of multidimensional Hamiltonian systems. Physica A, {\bf 390}, 3290-3307 (2011)

\bibitem{Arnold1967} Arnold, V.~I. and Avez, A.: {Probl\`{e}mes} {Ergodiques} de la {M\'{e}canique} {Classique}, Gauthier-Villars, Paris, 1967 \& Benjamin, New York, 1968, (1967)

\bibitem{Athan:2009a} Athanassoula, E., Romero-G\'{o}mez, M. and Masdemont, J.~J.: Rings and spirals in barred galaxies - I. Building blocks, Mon. Not. R. Astron. Soc., {\bf 394}, 67-81 (2009a)
\bibitem{Athan:2009b} Athanassoula, E., Romero-G\'{o}mez, M., Bosma, A., Masdemont,
    J.~J.: Rings and spirals in barred galaxies - II. Ring and spiral morphology , Mon. Not. R. Astron. Soc., {\bf 400}, 1706-1720 (2009b)
\bibitem{Athan:2010}  Athanassoula, E., Romero-G\'{o}mez, M., Bosma, A. and Masdemont,
    J.~J.: Rings and spirals in barred galaxies - III. Further comparisons and links to observations, Mon. Not. R. Astron. Soc., {\bf 407}, 1433-1448 (2010)

\bibitem{Baldovin2004a} Baldovin, F. and Brigatti, E. and Tsallis, C.: Quasistationary states in low-dimensional {Hamiltonian} systems,  Phys. Lett. A, {\bf 320}, 254-260 (2004)
\bibitem{Baldovin2004b} Baldovin, F. and Moyano, M.~G. and Majtey, A.~P. and Robledo, A. and Tsallis, C.: Ubiquity of metastable to stable crossover in weakly chaotic dynamical systems, Physica {\rm A}, {\bf 340}, 205-218 (2004)

\bibitem{Benettin1980a} Benettin, G., Galgani, L., Giorgilli, A. and Strelcyn, J.~M.: Lyapunov characteristic exponents for smooth dynamical systems and for {Hamiltonian} systems: {A} method for computing all of them. {Part} 1: {Theory}. Meccanica, {\bf 15}, 9-20 (1980a)
\bibitem{Benettin1980b} Benettin, G., Galgani, L., Giorgilli, A. and Strelcyn, J.~M.: Lyapunov characteristic exponents for smooth dynamical systems and for {Hamiltonian} systems: {A} method for computing all of them. {Part} 2: {Numerical} application. Meccanica, {\bf 15}, 21-30 (1980b)

\bibitem{Cachuchoetal:2010} Cachucho, F., Cincotta, P.~M. and Ferraz-Mello, S.: Chirikov diffusion in the asteroidal three-body resonance (5, -2, -2), Celest. Mech. Dyn. Astron., {\bf 108}, 35-58 (2010)

\bibitem{CinGiochap:2008} Cincotta, P.~M. and Giordano, C.~M.: Nonlinear Phenomena Research, Nova Science INC, Hauppauge, New York (2008)

\bibitem{Con_spr} Contopoulos, G.: Order and chaos in dynamical astronomy, Springer-Verlag, Berlin (2002)
\bibitem{ConHa:2008} Contopoulos, G., Harsoula, M., Stickiness in Chaos. Int. J. Bif. Chaos, {\bf 18}, 2929-2949 (2008)
\bibitem{ConHa:2010} Contopoulos, G. and Harsoula, M.: Stickiness effects in chaos, Celest. Mech. Dyn. Astron., {\bf 107}, 77-92 (2010)

\bibitem{Eckmann1985} Eckmann, J.~P. and Ruelle, D.: Ergodic theory of chaos and strange attractors, Rev. Mod. , 617-656, (1985)

\bibitem{Fer} Ferrers, N.~M.: Quart. J. Pure Appl. Math., {\bf 14}, 1 (1877)

\bibitem{Gio:2004} Giordano, C.~M. and Cincotta, P.~M.:	Chaotic diffusion of orbits in systems with divided phase space, Astron. Astrophys., {\bf 423}, 745-753 (2004)

\bibitem{HaKa:2009} Harsoula, M. and Kalapotharakos, C.: Orbital structure in N-body models of barred-spiral galaxies, Mon. Not. R. Astron. Soc., {\bf 394}, 1605-1619 (2009)

\bibitem{Hilhorst2010} Hilhorst, H.~J.: {Note} on a {$q$-modified} {Central} {Limit} {Theorem}, J. Stat. Mech., {\bf 10}, P10023 (2010)

\bibitem{KVC} Kalapotharakos, C., Voglis, N. and Contopoulos, G.: Chaos and secular evolution of triaxial N-body galactic models due to an imposed central mass Astron. Astrophys., {\bf 428}, 905-923 (2004)

\bibitem{KauCo} Kaufmann, D.~E. and Contopoulos, G.: Self-consistent models of barred spiral galaxies. Astron. Astrophys., {\bf 309}, 381-402 (1996)

\bibitem{katok1980} Katok, A.: Liapunov exponents, entropy and periodic orbits for diffeomorphisms, Publications Math\'{e}matiques de l'IH\'{E}S, {\bf 51}, 137-173 (1980)
	
\bibitem{KP11} Katsanikas, M. and Patsis, P.~A.: The structure of invariant tori in a 3D galactic potential. Int. J. Bif. Chaos, {\bf 21}, 467-496 (2011)

\bibitem{ManosPhD} Manos, T.: PhD Thesis, Universit\'e de Provence (Aix-Marseille I), France (2008)
\bibitem{Manos_etal:2008} Manos, T., Skokos, Ch., Athanassoula, E. and Bountis, T.: Studying the global dynamics of conservative dynamical systems using the SALI chaos detection method. Nonlin. Phenom. Compl. Syst., {\bf 11}, 171-176 (2008)
\bibitem{ManAthan:2009} Manos, T. and Athanassoula, E.: Dynamical study of 2D and 3D barred galaxy models. Eds Contopoulos G. and Patsis P., Chaos in Galaxies, Springer-Verlag, Berlin-Heidelberg (ASSP), p.~115-122 (2009)
\bibitem{MA11a} Manos, T. and  Athanassoula, E.: Regular and chaotic orbits in barsolid black galaxies - I. Applying the SALI/GALI method to explore their distribution in several models. Mon. Not. R. Astron. Soc., {\bf 415}, 629-642 (2011a)
\bibitem{MA11b} Manos, T. and  Athanassoula, E.: Regular and chaotic orbits in bar galaxies - II. Observable chaos. (in preparation) (2011b)

\bibitem{Miy:1975} Miyamoto, M. and Nagai, R.: Three-dimensional models for the distribution of mass in galaxies. Astron. Soc. of Japan, {\bf 27}(4), 533-543 (1975)

\bibitem{PAQ1997} Patsis, P.~A., Athanassoula, E. and Quillen, A.~C.:
	Orbits in the Bar of NGC 4314 Astron. Astrophys., {\bf 483}, 731-744 (1997)
\bibitem{PSA02} Patsis, P.~A., Skokos, Ch. and Athanassoula, E.: Orbital dynamics of three-dimensional bars - III. Boxy/peanut edge-on profiles. Mon. Not. R. Astron. Soc., {\bf 337}, 578-596 (2002)
\bibitem{PSA03a} Patsis, P.~A., Skokos, Ch. and Athanassoula, E.: Orbital dynamics of three-dimensional bars - IV. Boxy isophotes in face-on views. Mon. Not. R. Astron. Soc., {\bf 342}, 69-78 (2003a)
\bibitem{PSA03b} Patsis, P.~A., Skokos, Ch. and Athanassoula, E.: On the 3D dynamics and morphology of inner rings. Mon. Not. R. Astron. Soc., {\bf 346}, 1031-1040 (2003b)
\bibitem{Pats:2006} Patsis, P.~A.: The stellar dynamics of spiral arms in barred spiral galaxies
    Mon. Not. R. Astron. Soc., {\bf 369}, L56-L60 (2006)

\bibitem{Pesin1976} Pesin, Y.~B.: Invariant manifold families which correspond to nonvanishing characteristic exponents, Izv. Akad. Nauk. SSSR Ser. Mat., {\bf 40}, 1332-1379 (1976)

\bibitem{Pfe:1} Pfenniger, D.: The 3D dynamics of barsolid black galaxies. Astron. Astrophys., {\bf 134}, 373-386 (1984)

\bibitem{Plum} Plummer, H.~C.: On the problem of distribution in globular star clusters. Mon. Not. R. Astron. Soc., {\bf 71}, 460-470 (1911)

\bibitem{NuRe} Press, W.~H., Teukolsky, S.~A., Vetterling, W.~T. and Flannery, B.~P.,
    Numerical Recipes in Fortran 77: The Art of Scientific Computing, 2nd edn.
    Cambridge Univ. Press, Cambridge (1986)

\bibitem{Rice1995} Rice, J.: Mathematical statistics and data analysis, Duxbury Press (Second edition) (1995)

\bibitem{Rom:2006} Romero-G\'{o}mez, M., Masdemont, J.~J., Athanassoula, E. and
    Garc\'{i}a-G\'{o}mez, C.: The origin of rR1 ring structures in barred galaxies Astron. Astrophys., {\bf 453}, 39-45 (2006)
\bibitem{Rom:2007} Romero-G\'{o}mez, M., Athanassoula, E., Masdemont, J.~J.,   Garc\'{i}a-G\'{o}mez, C.: The formation of spiral arms and rings in barred galaxies  Astron. Astrophys., {\bf 472}, 63-75 (2007)

\bibitem{Ruelle1979} Ruelle, D.: Ergodic theory of differentiable dynamical systems, Phys. Math. IHES, {\bf 50}, 275-306, (1979)

\bibitem{Tsallisbook2009} Tsallis, C.: {Introduction} to {Nonextensive} {Statistical} {Mechanics}: {Approaching} a complex world, Springer, New York (2009)
\bibitem{TsallisTirnakli2010} Tsallis, C. and Tirnakli, U.: Nonadditive entropy and {Nonextensive Statistical Mechanics} - {Some} central concepts and recent applications, Journal of Physics: Conference Series, {\bf 201}, 012001 (2010)

\bibitem{Sinai1972} Sinai, Y. G.: Gibbs measures in ergodic theory, Uspekhi Matematicheskikh Nauk, {\bf 27}, 21-64 (1972)

\bibitem{Sk_sali:2001} Skokos, Ch.: Alignment indices: a new, simple method for determining the ordered or chaotic nature of orbits. J. Phys. A: Math. Gen., {\bf 34}, 10029-10043 (2001)
\bibitem{SPA02a} Skokos, Ch., Patsis, P.~A. and Athanassoula, E.: Orbital dynamics of three-dimensional bars - I. The backbone of three-dimensional bars. A fiducial case. Mon. Not. R. Astron. Soc., {\bf 333}, 847-860 (2002a)
\bibitem{SPA02b} Skokos, Ch., Patsis, P.~A. and Athanassoula, E.: Orbital dynamics of three-dimensional bars - II. Investigation of the parameter space. Mon. Not. R. Astron. Soc., {\bf 333}, 861-870 (2002b)
\bibitem{SBA_gali:2007} Skokos, Ch., Bountis, T. and Antonopoulos, Ch.: Geometrical properties of local dynamics in Hamiltonian systems: The Generalized Alignment Index (GALI) method. Physica D, {\bf 231}, 30-54 (2007)
\bibitem{Skokos2010} Skokos, Ch.: The {Lyapunov} characteristic exponents and their computation. Lecture Notes in Physics, {\bf 790}, 63-135, (2010)

\end{thebibliography}
\end{document}